\pdfoutput=1
\documentclass[conference]{IEEEtran}
\IEEEoverridecommandlockouts
\usepackage[noadjust]{cite}
\usepackage{amsmath,amssymb,amsfonts}
\usepackage{algorithmic}
\usepackage{graphicx}
\usepackage[T1]{fontenc}
\usepackage{float}
\usepackage{textcomp}

\def\BibTeX{{\rm B\kern-.05em{\sc i\kern-.025em b}\kern-.08em
   T\kern-.1667em\lower.7ex\hbox{E}\kern-.125emX}}
    
\begin{document}
%
\title{Multi-level Attention Model for Weakly Supervised Audio Classification}

\author{\IEEEauthorblockN{Changsong Yu\IEEEauthorrefmark{1},
Karim Said Barsim\IEEEauthorrefmark{1}, Qiuqiang Kong\IEEEauthorrefmark{2} and
Bin Yang\IEEEauthorrefmark{1}}
\IEEEauthorblockA{\IEEEauthorrefmark{1} Institute of Signal Processing and System Theory, University of Stuttgart, Germany\\
\IEEEauthorrefmark{2}Center for Vision, Speech and Signal Processing, University of Surrey, UK}}
\maketitle

\maketitle

\begin{abstract}
In this paper, we propose a multi-level attention model to solve the weakly labelled audio classification problem. The objective of audio classification is to predict the presence or absence of audio events in an audio clip. Recently, Google published a large scale weakly labelled dataset called Audio Set, where each audio clip contains only the presence or absence of the audio events, without the onset and offset time of the audio events. Our multi-level attention model is an extension to the previously proposed single-level attention model. It consists of several attention modules applied on intermediate neural network layers. The outputs of these attention modules are concatenated to a vector followed by a multi-label classifier to make the final prediction of each class. Experiments shown that our model achieves a mean average precision (mAP) of 0.360, outperforms the state-of-the-art single-level attention model of 0.327 and Google baseline of 0.314. 

\end{abstract}
\IEEEpeerreviewmaketitle
\section{Introduction}

Audio classification aims to predict the presence or absence of audio events in an audio clip. It has attracted many interests in recent years and has many applications such as multimedia information retrieval and public surveillance \cite{mesaros2017dcase, mesaros2016tut}. Before 2017, many of the audio datasets are relatively small compared with large image dataset such as ImageNet \cite{deng2009imagenet}. For example, urban sound dataset \cite{salamon2014dataset} contains about 27 hours of urban sound records with 3075 samples. ESC-50 dataset \cite{piczak2015esc} consists of 2000 environmental recordings across 50 classes. The detection and classification of acoustic scenes and events (DCASE) challenge 2013, 2016, 2017 \cite{mesaros2017dcase, mesaros2016tut, stowell2015detection} datasets are comprised of several hours of data. Recently, Google published a large scale audio classification dataset called Audio Set \cite{gemmeke2017audio} consisting of 5,800 hours two million human-labeled 10-second audio clips covering 527 audio categories. 

In Audio Set, each audio clip contains one or several labels, such as ``cat'', ``speech'' and ``park'' \cite{gemmeke2017audio}. Audio Set is a \textit{weakly labelled dataset}, that is, only the presence or absence of audio events are known in an audio clip, without knowing the onset and offset time of the audio events. In the weakly labelled dataset, the duration of the audio events varies depending on audio categories. Some audio events in an audio clip last for several seconds such as ``speech'', while some audio events only last for hundreds of milliseconds, such as ``gunshot''. 

To solve the weakly labelled data problem, many methods have been proposed such as multiple instance learning (MIL) \cite{maron1998framework} and has been applied on weakly labelled audio classification \cite{kumar2016audio}. Later, Kong et al. \cite{kong2017audio} proposed an single-level attention model for audio classification and outperforms both the MIL method \cite{kumar2016audio} and the Google baseline deep neural network system \cite{gemmeke2017audio} on Audio Set classification. This model consists of three fully connected layers followed by one attention module. The motivation of the attention module is that different segments in an audio clip contributes differently to the classification of the audio clip. For example, the segments containing an event should be attended and the segments containing irrelevant noise should be ignored. 

However, the shortcoming of the single-level attention model is that substantial information from the intermediate neural network layers (the three fully connected layers) is disregarded. Many works \cite{lee2017multi,meng2017multi, roth2015deeporgan} explored that features from intermediate neural network layers contains rich information. Lee et. al. \cite {lee2017multi} explored that the audio classification performance can be improved by concatenating features from different intermediate neural network layers. In addition, using multi-level features has been found to be effective not only for audio tasks, but also for vision tasks. Meng et al. \cite{meng2017multi} extracted features from different layers of a deep CNN. These features are concatenated to a representation and significantly outperforms the non-concatenated features \cite{meng2017multi}. 

Inspired by the success of multi-level representation \cite{lee2017multi, meng2017multi}, we expand Kong's model \cite {kong2017audio} to a \textit{multi-level attention model}. Firstly, attention modules are used on the intermediate neural network layers. Then, the outputs of the attention modules are concatenated to a vector. Finally, a fully connected layer with sigmoid activation function is utilized to predict the presence probability of each class. 

The paper is organized as follows. Section II introduces some related works. Section III introduces the single-level attention model \cite {kong2017audio}. Section IV describes the proposed multi-level attention module. Section V shows experimental results. Section VI concludes and forecasts future work. 

\section{Related Works}

\textbf{Audio classification}: Audio classification has attracted many attention in recent years. Some representative challenges including DCASE 2013 \cite{stowell2015detection}, DCASE 2016 \cite{mesaros2016tut} and DCASE 2017 \cite{mesaros2017dcase}. Hidden Markov models have been used to model audio events in \cite{peng2009healthcare}. Non negative matrix based methods were applied to learn the dictionary of audio events \cite{bisot2016supervised}. Recently, neural network based methods including fully connected neural networks \cite{kong2016deep}, convolutional neural networks (CNN) \cite{choi2016automatic} have been applied on audio classification and achieved the state-of-the-art performance. 

\textbf{Attention module}: The concept of attention module is first introduced in natural language processing \cite{bahdanau2014neural}. Attention module allows deep neural networks to focus on relevant instances and ignore irrelevant instances in a bag. It has been successfully applied in machine translation \cite{bahdanau2014neural}, face detection \cite{sharma2015action}, image classification \cite{shih2016look} and captioning \cite{xu2015show}. It is also utilized in the domain audio classification \cite {xu2017large}. 

\section{Dataset}
Audio Set \cite{gemmeke2017audio} consists of over two million samples. There are 527 classes in the current version. Audio Set is a multi-label dataset and each audio clip has one or several labels. Google created Audio Set through transfer learning. In the pre-training stage, two billion 10-second audio clips from YouTube covering more than 30,000 classes are collected and called YouTube 100M \cite{hershey2017cnn}. Log Mel spectrogram with size of $ 96 \times 64 $ along time and frequency axis is extracted as feature for each audio clip. Then, a ResNet-50 model is trained using this YouTube 100M data. This trained ResNet-50 is later used as a feature extractor. After the pre-training stage, two million 10-second audio clips covering 527 classes are collected. The log Mel spectrogram of each audio clip is presented to the trained ResNet-50 model to extract the bottleneck features. In this process, each audio clip is compressed into 10 bottleneck features which we call the collection of these 10 features as a sample. Each feature has a dimension of 128. These two million samples constitute Audio Set.

\section{Single-level attention model}\label{single-level}
In this section, we will introduce the single-level attention model proposed in \cite{kong2017audio}. 

To illustrate the notation, let $x_{t},\, t = 1, 2, ..., T$ be the t-th bottleneck feature with a dimension $M=128$. Each sample in Audio Set has $T=10$ bottleneck features. $ K = 527 $ is the number of classes.

In the single-level attention model, each bottleneck feature $ x_{t} $ is presented to a trainable embedding mapping $ f_{emb}(\cdot) $ to extract an embedded feature $ h_{t} $:

\begin{equation}
h_{t} = f_{emb}(x_{t})
\tag{1} 
\label{eq:1}
\end{equation}

Furthermore, an attention module is applied on the $T$ embedded features to attain the class probabilities for the input sample:

\begin{equation}
y(\mathbf{h}) = \frac{1}{\sum_{t=1}^{T}v(h_{t})} \sum_{t=1}^{T}v(h_{t})f(h_{t})
\tag{2} 
\label{eq:2}
\end{equation}

\begin{figure}[h!]
\renewcommand\thefigure{1}
  \centering
  \includegraphics[keepaspectratio, width=0.9\columnwidth]{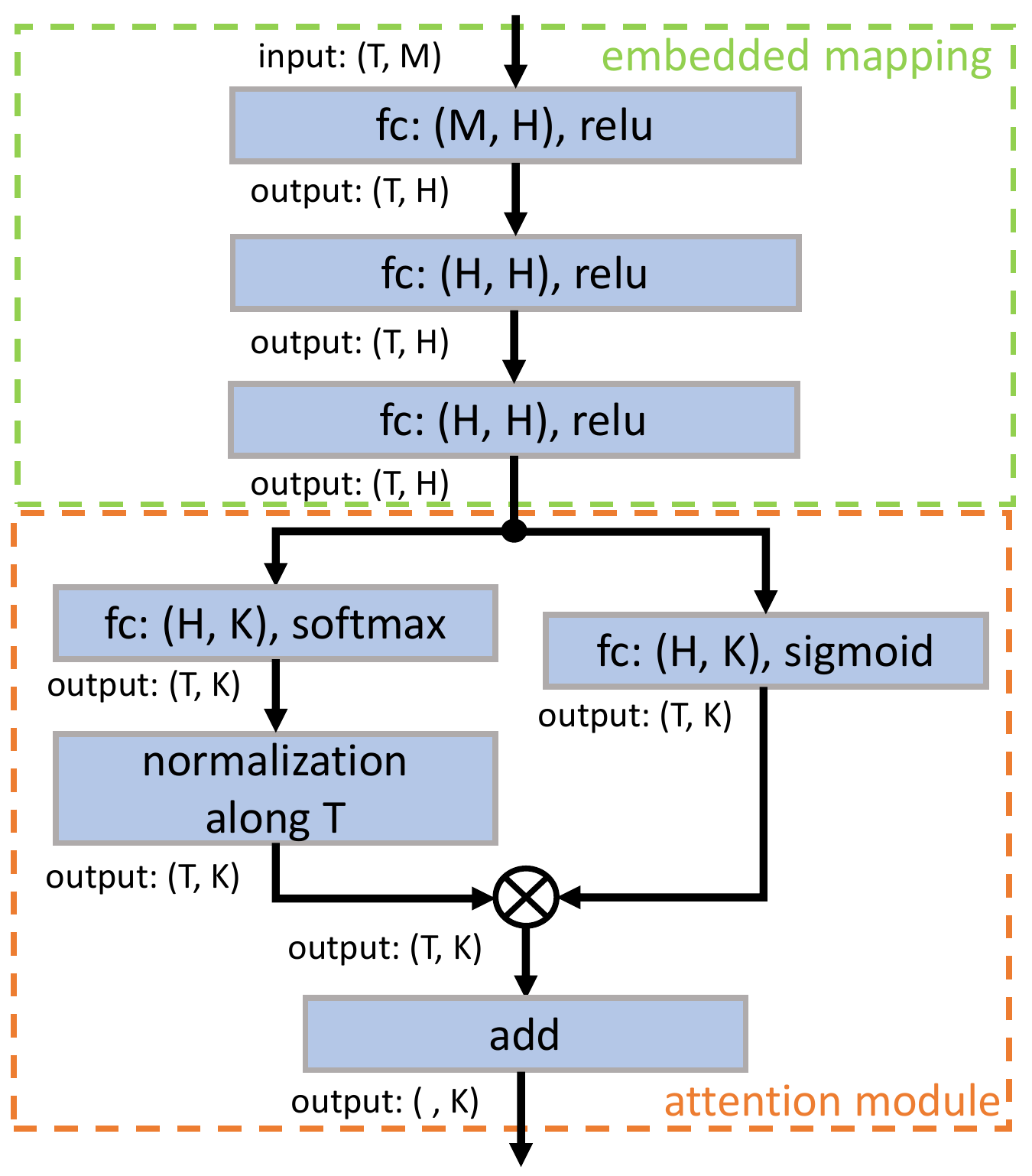}
  \caption{Architecture of the single-level attention model \cite {kong2017audio}}
  \label{fig:1}
\end{figure}

\noindent where $ \mathbf{h} = \left [ h_{1}, ..., h_{T} \right ] $ is the concatenation of the embedded features. Non-negative function $v(\cdot)$ determines how much an embedded feature $ h_{t} $ should be attended or ignored and $ f(\cdot) $ denotes the classification output on an embedded feature $ h_{t} $. The attention module has ability to ignore irrelevant sound segments such as background noise and silences, and attend to the sound segments with audio events. 

The implementation of the single-level attention model is shown in Fig. 1. The first part is an embedded mapping $ f_{emb}(\cdot) $ modeled by three fully connected neural layers with $H$ units. The second part is an attention module described by Equation (\ref{eq:2}). The attention non-negative mapping $ v_{k}(\cdot) $ and the classification mapping $ f_{k}(\cdot) $ are modeled by a softmax function and sigmoid function, respectively. The normalization applied after $ v_{k}(\cdot) $ ensures the attention is normalized. Finally, the prediction is obtained by element-wise multiplication of the classification output and normalized attention output.

\section{Multi-level attention model}

Many works have explored that using multi-level features from intermediate layers of neural networks can promote the audio or image classification performance \cite{lee2017multi, meng2017multi}. We propose to extend the single-level attention model in Section \ref{single-level} to multi-level attention model in our paper.

The architecture of the proposed multi-level attention model is shown in Fig. 2. Instead of applying a single-level attention model after the fully connected neural network, multiple attention modules are applied after intermediate layers as well. These attention modules aim to capture different level information. We denote the feedforward mappings as $ g_{l}(\cdot) $ and the activations of the intermediate layers as $ h^{(l)} $, where $l$ is the number of embedded mappings. The feed-forward neural network can be written as:

\begin{equation}
  \begin{cases}
               h_{t}^{(1)} = g_{1}(x_{t}) &  \\
               h_{t}^{(l)} = g_{l}(h_{t}^{(l-1)}) & l = 2,3,...,L\\
  \end{cases} 
  \tag{3} \label{eq:3}
\end{equation}

\noindent where each forward mapping $ g_{l}(\cdot) $ may consists of several fully connected layers in series (Fig. 2). For the single-level attention model, the prediction is produced by $ y^{(L)} = y(\mathbf{h}^{L}) $ follows Equation (\ref{eq:2}) where $ \mathbf{h}^{(L)}=\left [ h_{1}^{(L)}, ..., h_{T}^{(L)} \right ] $. 

In the proposed multi-level attention model, each $ l $-th attention module produces a prediction $ y^{(l)}=y(\mathbf{h}^{(l)})$. Each prediction $ y^{(l)} \in \left [ 0, 1 \right ] ^K $. Then, all the predictions are concatenated to a vector $ u \in \left [ 0, 1 \right ] ^{KL} $:

\begin{equation}
u=\left [ y^{(1)}, ..., y^{(L)} \right ] \tag{4}\label{eq:4}
\end{equation}

\begin{figure}[h!]
\renewcommand\thefigure{2}
  \centering
  \includegraphics[keepaspectratio, width=0.33\textwidth]{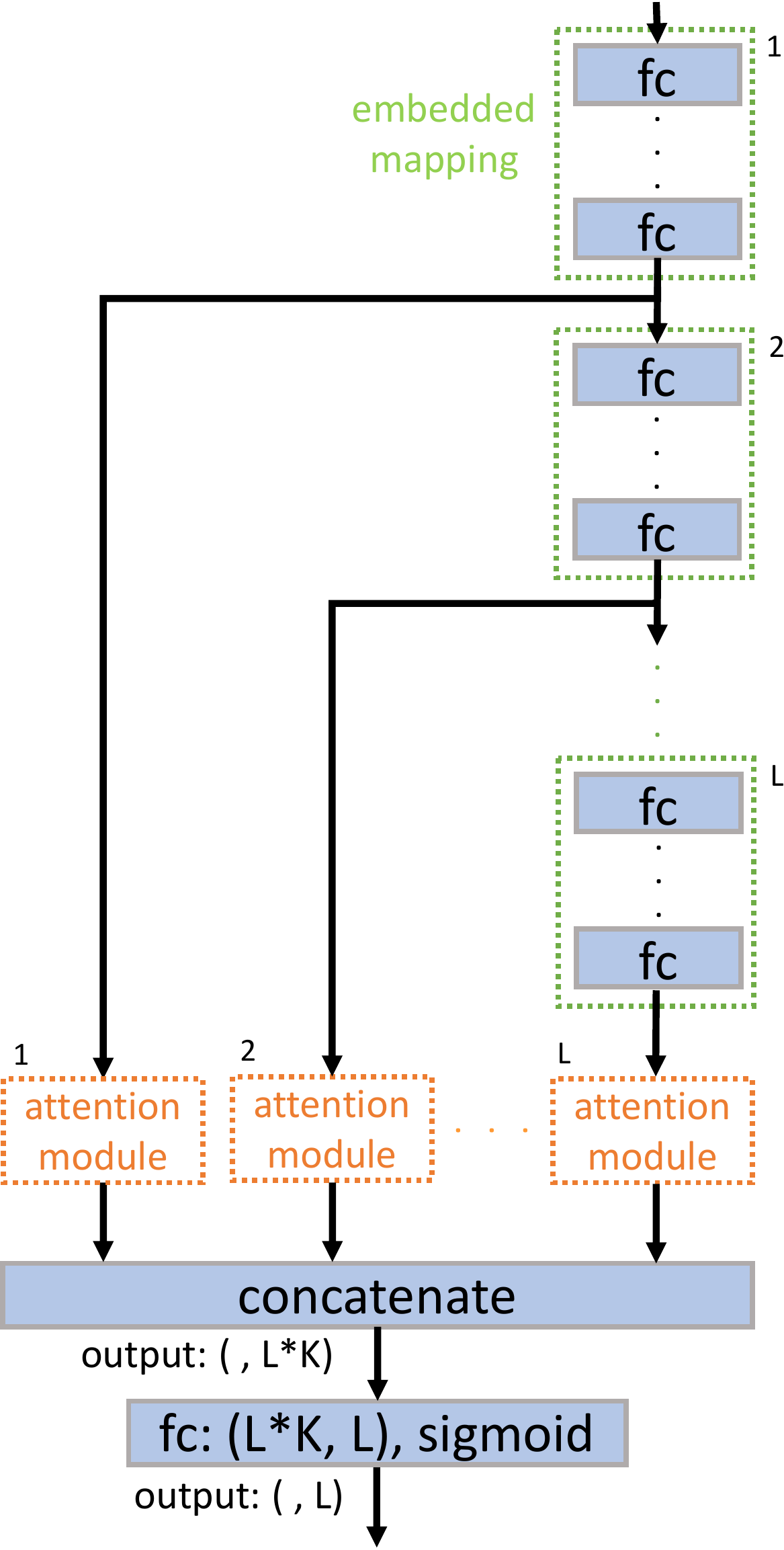}
  \caption{Architecture of the multi-level attention model}
  \label{fig:2}
\end{figure}

Finally, a fully connected layer followed by sigmoid non-linearity is applied on the concatenated vector $ u $ to attain the class probabilities $ z \in \left [ 0, 1 \right ] ^K$ of the audio classes. 

\begin{equation}
z = \phi(Wu + b) \tag{5} \label{eq:5}
\end{equation}

\noindent where the $ W \in \mathbb{R}^{KL \times K} $ and $ b \in \mathbb{R}^{K} $ represent the weight matrix and the bias, separately. Sigmoid non-linearity $\phi(\cdot)$ is used for multi-label classification. 

\section{Experiments}
\begin{figure*}[h!]
\includegraphics[keepaspectratio,width=1\textwidth]{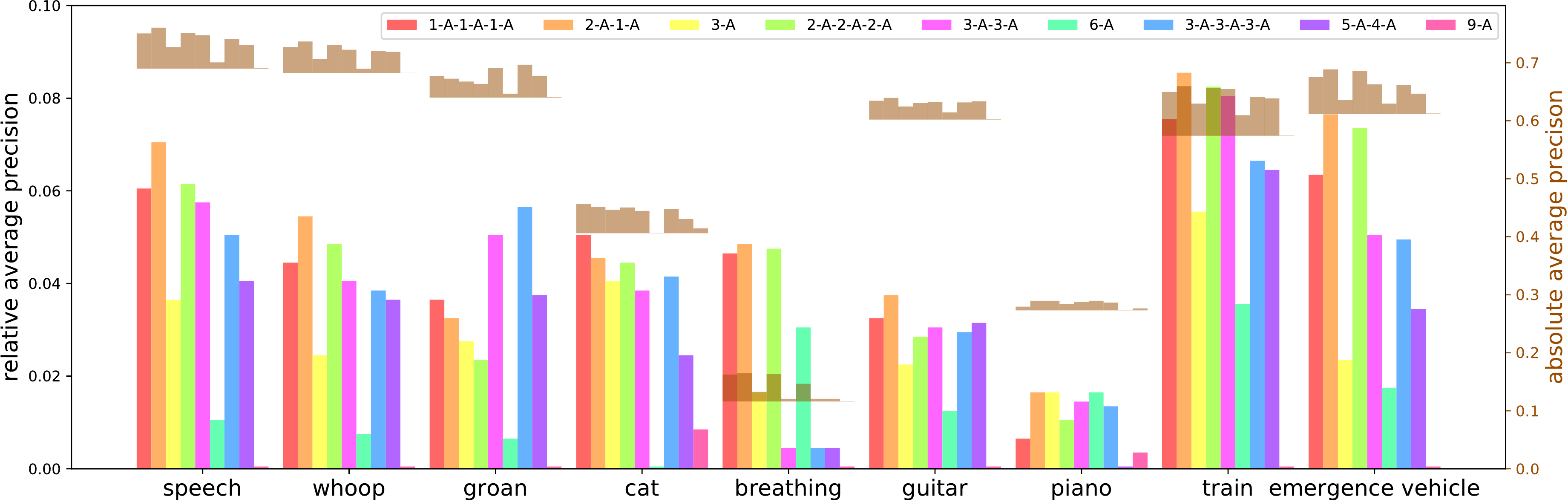}
\label{fig:imagea}
\caption{ Average precision (AP) results of all single-level or multi-level attention models for nine randomly selected classes. The left black bar-graph scaled by the y-axis on the left-side represents the relative AP to the lowest AP among all models on a class. For example, the lowest AP among all models of the class "speech" is the AP of 9-A. The relative AP of 9-A of this class is 0 and that of 5-A-4-A is 0.04. The right brown bar-graph scaled by the y-axis on the right side represents the absolute AP. For example, the APs of 5-A-4-A and 9-A for the class "speech" are 0.730 and 0.690, separately.}
\end{figure*}

\subsection{Training details}
In order to comprehensively compare the performance of single-level and multi-level attention models, we implemented nine variants of single- (3-A, 6-A, 9-A) or multi-level attention models (1-A-1-A-1-A, 2-A-1-A, 2-A-2-A-2-A, 3-A-3-A, 3-A-3-A-3-A, 5-A-4-A) which are shown in Table I. The model 2-A-1-A represents two attention modules are applied after the 2nd and 3rd fully connected layers. The model 2-A-2-A-2-A represents three attention modules are applied after the 2nd, 4th and 6th fully connected layers. Each fully connected layer in all embedded mappings consists of 600 hidden units followed by ReLU activation function \cite{nair2010rectified}. Dropout is used to prevent overfitting \cite{srivastava2014dropout} with dropout rate of 0.4. Batch normalization \cite{ioffe2015batch} is applied to speed up training and prevent overfitting. The weights and biases are default-initialized in Keras 2.0.8. Adam optimizer \cite{kingma2014adam} with learning rate of 0.001 is used. Batch size is set to 500. The setting of these hyper-parameters follows the configuration in \cite{kong2017audio}. Code has been made publicly available here \footnote[1]{\url{https://github.com/ChangsongYu/Eusipco2018\textunderscore Google\textunderscore AudioSet.git}}.

\subsection{Evaluation Metrics}
To evaluate our model, we use three metrics of the Google's benchmark: mean average precision (mAP), area under curve (AUC) and d-prime.   
The mAP is the mean of average precision over all classes. The mAP is calculated by:

\begin{equation}
    mAP = \frac{1}{K}\sum_{c=1}^{K}\sum_{n=1}^{N}p_{c,n}\Delta{r_{c,n}} \tag{6} \label{eq:6}
\end{equation}

where $p_{\textbf{c,n}}$ is the precision at $n$-th positive sample of $c$-th class. $N$ is the number of positive samples for each class. $\Delta{r_{c,n}}$ is equal to $\frac{1}{N}$. 

The AUC is area under the true positive-false positive rate curve. True positive rate (TPR) is a probability of correctly classifying a positive sample. False negative rate (FPR) is a probability of incorrectly classifying a negative sample as positive.

Google found that AUC has awkward scaling when more than 0.9. Hence, another metric d-prime which is a simple deterministic function of AUC is applied. The d-prime is the separation of the means between two unit-variance Gaussians. We assume that the score distributions of the positive and negative samples are two unit-variance Gaussians. Then, we can calculate the d-prime directly from AUC. Its equation shows below:

\begin{equation}
d{\text -}prime = \sqrt{2}F_{x}^{-1}(AUC) \tag{7} \label{eq:7}
\end{equation}

$F_{x}^{-1}$ is inverse of the cumulative distribution function and defined by: 
\begin{equation}
    F_{x}(x) = \int_{-\infty}^{x}\frac{1}{\sqrt{2\pi}}e^{\frac{-({x-\mu})^2}{2}}dx  \tag{8} \label{eq:8}
\end{equation}

The larger the values of all three metrics, the better the audio classification.

\subsection{Analysis}

The first two rows in Table I show the results of Google's benchmark \cite{gemmeke2017audio} without attention model and Kong's result of the single-level attention model \cite{kong2017audio}. All of multi-level attention models outperform Google's baseline and single-level attention model in all of mAP, AUC, and d-prime. The best multi-level attention model is 2-A-1-A when using two attention modules on the 2nd and 3rd intermediate layers, where mAP of 0.360 is achieved, outperforms 0.327 in single-level attention model \cite{kong2017audio} and 0.314 of Google's baseline system. The reason for the good performance using multi-level attention model is that the multi-level features extracted from the intermediate layers provide various representations, and then each attention module can filter the unrelated information of each feature. In addition, different classes may favor different layer of features and the last fully connected layer of each multi-level attention model can automatically select best feature for each class by the weight parameters. 

When comparing all variants of the single-level attention model (3-A, 6-A, 9-A), it was observed that the performance
notably degrades as the number of fully connected layers is increased. This attribute to that the feature extracted from a deep fully connected layer (e.g. 6th and 9th fully connected layer) has worse discriminative power than that of a shallow layer (e.g. 3rd fully connected layer).


\begin{table}[htbp]
\label{result}
\caption{Comparisons of results of multi-level attention model}
\begin{center}
\begin{tabular}{*4l}
\textbf{Model } & \textbf{mAP} &\textbf{AUC} & \textbf{d-prime} \\
\hline
Benchmark & 0.314 & 0.9590 & 2.452 \\
\hline
Kong \cite{kong2017audio} & 0.327 & 0.9650 & 2.558 \\
\hline
1-A-1-A-1-A & 0.357 & 0.9693 & 2.645 \\
\hline
\textbf{2-A-1-A} & \textbf{0.360} & \textbf{0.9700} & \textbf{2.660} \\
\hline
3-A & 0.336 & 0.9668 & 2.596 \\
\hline
2-A-2-A-2-A & 0.358 & 0.9695 & 2.650 \\
\hline
3-A-3-A & 0.355 & 0.9690 & 2.639 \\
\hline
6-A & 0.311 & 0.9571 & 2.430 \\
\hline
3-A-3-A-3-A & 0.353 & 0.9687 & 2.633 \\
\hline
5-A-4-A & 0.340 & 0.9676 & 2.612 \\
\hline
9-A & 0.305 & 0.9388 & 2.185 \\
\hline
\end{tabular}
\end{center}
\end{table}

\subsection{Performance visualization of individual classes}

In addition, we investigate all variants of our single-level or multi-level attention model by comparing average precision (AP) of nine randomly selected classes are shown in Figure 3. For each class, the color bars plotted below is the relative improvement of AP and the bars plotted above is the absolute AP. The APs of classes such as speech and whoop are close to 0.7. In contrast, APs of many classes such as breathing are lower than 0.2.

Figure 3 shows that the multi-level attention models don't always achieve better performance on all classes than the single-level attention models. For the class "piano", the model 6-A outperforms the models 2-A-2-A-2-A and 3-A-3-A. We also observe that different classes favor different models. For example, the classes "speech", "whoop", "breathing", "guitar", "train" and "emergence vehicle" favor the model 2-A-1-A. However, the class "groan" favors the model 3-A-3-A-3-A. Overall, we can ensure that the performance of classification consistently increases on most classes when the multi-level features are concatenated and 2-A-1-A is the best architecture.       

\section{Conclusion}
In this work, we introduced a multi-level attention model in addressing weakly labelled audio classification problem on Audio Set. The experimental results showed the effectiveness of concatenating multi-level features. The best result of our multi-level attention model successfully exceeds the Google's benchmark and the previous state of the art results. In future work, we plan to combine multi-scale and multi-level features together to train Audio Set.
\newpage
\nocite{*}
\bibliographystyle{IEEEtran}

\bibliography{eusipco_2018_multi_level_attention_model}

\end{document}